\documentclass[12pt]{iopart}

\usepackage{iopams}
\usepackage{hyperref}
\usepackage{graphicx}

\begin{document}

\article[Deep space experiment to measure $G$]{}{Deep space experiment to measure $G$}

\author{Michael R Feldman$^1$, John D Anderson$^2$\footnote{Retired.}, 
Gerald Schubert$^3$, Virginia Trimble$^4$, Sergei M Kopeikin$^5$ and Claus L\"{a}mmerzahl$^6$}

\address{$^1$ m-{}-y laboratories inc., West Hollywood, California 90046, USA}
\address{$^2$ Jet Propulsion Laboratory, California Institute of Technology, Pasadena, California 91109, USA}
\address{$^3$ Department of Earth, Planetary and Space Sciences, University of California, Los Angeles, California 90095, USA}
\address{$^4$ Department of Physics and Astronomy, University of California, Irvine, California 92697, USA}
\address{$^5$ Department of Physics and Astronomy, University of Missouri, Columbia, Missouri 65211, USA}
\address{$^6$ ZARM, University of Bremen, 28359 Bremen, Germany}

\ead{mrf@m--y.us, jdalya2@gmail.com, schubert@ucla.edu, vtrimble@astro.umd.edu, kopeikins@missouri.edu, claus.laemmerzahl@zarm.uni-bremen.de}

\begin{abstract}
Responding to calls from the National Science Foundation (NSF) for new proposals to measure the gravitational constant $G$, we offer an interesting experiment in deep space employing the classic gravity train mechanism. Our setup requires three bodies: a larger layered solid sphere with a cylindrical hole through its center, a much smaller retroreflector which will undergo harmonic motion within the hole and a host spacecraft with laser ranging capabilities to measure round trip light-times to the retroreflector but ultimately separated a significant distance away from the sphere-retroreflector apparatus. Measurements of the period of oscillation of the retroreflector in terms of host spacecraft clock time using existing technology could give determinations of $G$ nearly three orders of magnitude more accurate than current measurements here on Earth. However, significant engineering advances in the release mechanism of the apparatus from the host spacecraft will likely be necessary. Issues with regard to the stability of the system are briefly addressed.
\end{abstract}

\pacs{04.80.-y, 02.60.Cb}

\vspace{2pc}
\noindent{\it Keywords}: Experimental studies of gravity, gravitational constant, gravity train, numerical simulation

\submitto{\CQG}
\maketitle
%
%

\section{Introduction}

The ongoing situation with regard to inconsistent values for the gravitational constant $G$ \cite{Speake2014, Anderson2015, Schlamminger2015} has highlighted the necessity of more measurements and alternative approaches. Clear evidence of this comes from the NSF's recent solicitation on the matter (\href{http://www.nsf.gov/pubs/2016/nsf16520/nsf16520.htm}{Ideas Lab: Measuring ``Big G" Challenge (16-520)}). Previous approaches to measuring $G$ have relied heavily upon variations of Cavendish's original torsion balance scheme \cite{Cavendish1798}, accompanied with recent forays into atom interferometry \cite{Rosi2014}. A major issue with these experiments is the fact that each occurs in a laboratory on Earth, where one must combat Earth-based influences to accurately determine $G$. Nevertheless, the beginning of humanity's direct exploration of space over the last sixty years provides us with ample opportunities in fundamental physics, especially with regard to measuring weaker interactions such as gravitation. This comes from our ability to launch man-made probes on escape orbits out of the Solar System with deep space ($>25$~AU) as an ideal laboratory to remove the constraints imposed by Earth-based experiments.

Probing fundamental physics in space is not a new idea, e.g. \cite{Everitt2015, Amaro-Seoane2013, Cacciapuoti2011}. Precise tracking with the NASA Deep Space Network (DSN) has made it possible to determine the trajectories of spacecrafts to unprecedented accuracy, significantly improving our understanding of the motion of celestial bodies \cite{Folkner2014, Pitjeva2014, Fienga2011} and ushering in ever more stringent tests of general relativity \cite{Bertotti2003}. Suggestions that the DSN can be used to test fundamental physics have come recently in the form of dedicated mission proposals \cite{Dittus2005, Johann2008, Christophe2009, Wolf2009, Aguilera2014, Buscaino2015} with a crucial emphasis on constraining any potential outstanding deviations from general relativity \cite{Anderson2010, Iorio2015}. Our purpose with this article is to build upon these proposals by using deep space as our laboratory for an experiment to measure $G$.

\section{Methods}

The goals for a successful measurement of Newton's gravitational constant are clear given the seemingly periodic nature of inconsistent values here on Earth \cite{Anderson2015, Schlamminger2015}. One should not only have a measurement with direct dependence on the value of $G$ but also an experiment that can theoretically produce near-continuous measurements over a substantial timeframe with an apparatus subject to a minute and accountable number of non-gravitational forces. While stringent constraints on the time-variation of the solar gravitational mass parameter \cite{Folkner2014, Pitjeva2014, Fienga2011, Zhu2015, Iorio2016} make it very unlikely the actual value of $G$ is varying at the levels indicated in the aforementioned works, testing with an apparatus much smaller than the size of the Solar System is still an important step in constraining any time-dependence if scale-dependent effects were to exist \cite{Will2014}.

To achieve these goals, we suggest using the classic gravity train mechanism of a small body exhibiting simple harmonic motion through a tunnel extending along the diameter of a much more massive solid sphere. Traditionally, this mechanism has been discussed as a pedagogically useful thought experiment of how one might travel between locations on Earth \cite{Cooper1966, Kirmser1966, Klotz2015}. But, of course, Earth is not special in these examples. For an observer traversing through a spherically symmetric body of constant density having mass $M$ and radius $R$, one finds, according to Newtonian gravitation, that the observer oscillates between the ends of the tunnel with period
\begin{equation}
\label{SHOperiod}
T = 2\pi \sqrt{\frac{R^3}{MG}},
\end{equation}
assuming the observer has insignificant mass relative to that of the sphere.

\begin{figure}
\includegraphics{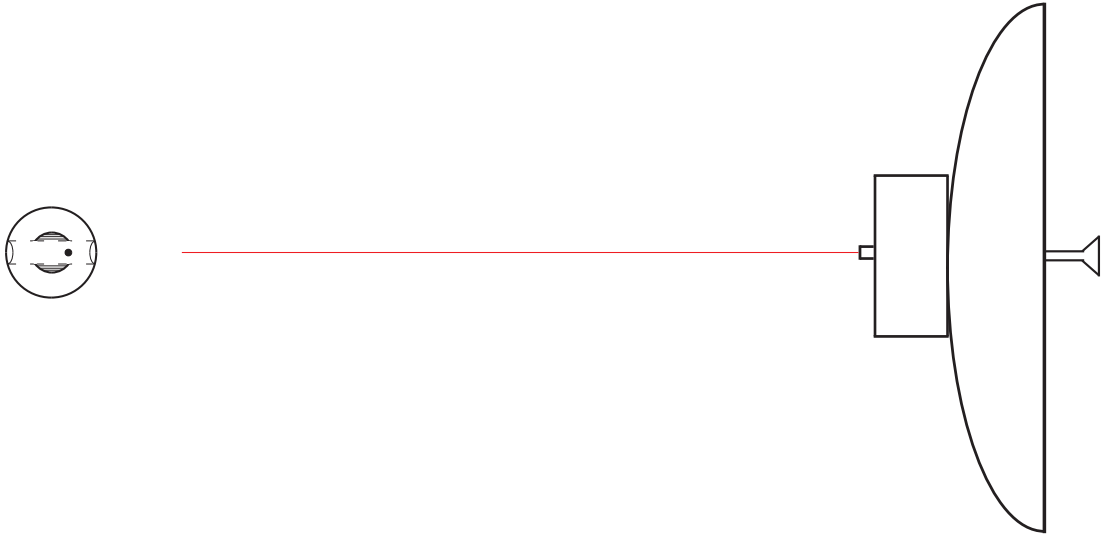}
\caption{Schematic of our proposed experimental setup. The smaller retroreflector exhibiting harmonic motion within the tunnel of the larger mass is represented by the black circle (on the left). Determinations of the round trip light-time from the host spacecraft (on the right) to the retroreflector using an onboard ranging system provide measurements of the period of the gravity train oscillator.}
\label{fig.1}
\end{figure}

Exploiting this periodic behavior for a determination of $G$, we offer the apparatus of \fref{fig.1}. Note that our oscillating ``observer" or ``train" is a small retroreflector meant to return guided range pulses from a host spacecraft for measurements of the round trip light-time to the retroreflector as determined by a stable clock onboard the host. Stored measurement data will eventually be downlinked through the DSN complex using the host spacecraft's radio communications with Earth. Ultimately, one can use these light-time measurements to build a profile of the oscillator's position in terms of host spacecraft clock time and thereby determine its period for each harmonic cycle within the larger mass. Since the period of the oscillator is explicitly dependent upon the value of $G$ and the values of $M$ and $R$ for the larger sphere will be known to a certain accuracy, we have the ability to obtain multiple determinations of $G$ over the course of our three body setup's radial trajectory out of the Solar System. Furthermore, as these measurements will be performed in the relative vacuum of deep space, external forces on the apparatus are accountable.

\begin{figure}
\includegraphics{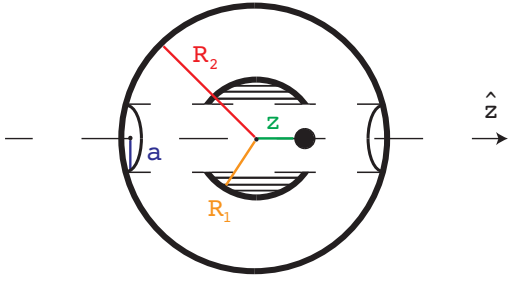}
\caption{Close-up of our spherical mass. The striped region marks a dense inner core.}
\label{fig.2}
\end{figure}

Nevertheless, equation~\eref{SHOperiod} results from overly simplistic assumptions. One can verify issues with stability exist for the constant density sphere with a cylindrical hole due to the projection of the gravitational field in the plane perpendicular to the symmetry axis being directed away from the axis for the majority of the hole. To combat this and maintain stability over the lifetime of the experiment, we suggest our larger sphere be composed of two layers: a dense uniform spherical inner core surrounded by a less dense, yet still uniform, spherical outer envelope (\fref{fig.2}). Taking into account the cylindrical hole of radius $a$, inner core of radius $R_1$ and density $\mu_1$, and outer envelope of radius $R_2$ and density $\mu_2$, we have for the Newtonian gravitational potential due to the two-layer sphere inside the hole (or, more adequately, the ``spherical ring"),
\begin{eqnarray}
\fl \Psi_g(\varrho,z) = \psi_1(\mu_1 - \mu_2, R_1,\varrho,z) + \psi_2(\mu_2, R_2,\varrho,z); \qquad \varrho < a,
\\ \fl \psi_i(\mu,R_i,\varrho,z) = A_i (z) + B_i(z)\varrho^2 +\mathcal{O}(\varrho^4), \label{gravPotentialLowOrder}
\\ \fl A_i(z) = G \pi \mu \bigg\{ \frac{(2R_i^2 - z\sqrt{R_i^2-a^2}-z^2)(R_i^2-2z\sqrt{R_i^2-a^2} +z^2)^{1/2}}{3z} \nonumber
\\ - \frac{(2R_i^2 + z\sqrt{R_i^2-a^2}-z^2)(R_i^2+2z\sqrt{R_i^2-a^2} +z^2)^{1/2}}{3z} \nonumber
\\ \fl + R_i^2\bigg[\mathrm{arctanh}\bigg(\frac{(R_i^2-2z\sqrt{R_i^2-a^2} +z^2)^{1/2}}{\sqrt{R_i^2-a^2}-z} \bigg) + \mathrm{arctanh}\bigg(\frac{(R_i^2+2z\sqrt{R_i^2-a^2}+z^2)^{1/2}}{\sqrt{R_i^2-a^2}+z}\bigg) \bigg] \nonumber
\\ + (R_i^2-a^2) \ln \bigg[\frac{-\sqrt{R_i^2-a^2} + z + (R_i^2-2z\sqrt{R_i^2-a^2} +z^2)^{1/2}}{\sqrt{R_i^2-a^2} + z + (R_i^2+2z\sqrt{R_i^2-a^2} + z^2)^{1/2}} \bigg]\bigg\},
\\ \fl B_i(z) = - \frac{G\pi \mu}{6z^3}\bigg[\frac{2R_i^4 -2R_i^2 z\sqrt{R_i^2-a^2}+z^2(a^2+z\sqrt{R_i^2-a^2} - z^2)}{(R_i^2-2z\sqrt{R_i^2-a^2} + z^2)^{1/2}} \nonumber
\\ - \frac{2R_i^4 + 2R_i^2 z\sqrt{R_i^2-a^2}+ z^2(a^2 - z\sqrt{R_i^2-a^2} - z^2)}{(R_i^2+2z\sqrt{R_i^2-a^2} + z^2)^{1/2}}  \bigg],
\end{eqnarray}
in powers of distance away from the symmetry axis, $\varrho$, from expanding the Green's function in cylindrical coordinates. Standard orbit fitting techniques can then be used to infer a value for $G$.

\begin{table}[b]
\caption{\label{tab.1}
Preliminary error analysis for our $G$ determination. $R$ is the radius of a near perfect sphere, $M$ is its total mass and $T$ is the period of the oscillator. The uncertainty in measuring $G$ relies upon the assumption of a sphere outer radius of $4.7$~cm and a laser ranging accuracy of $1.1$~nm.
}
\begin{indented}
\item[]\begin{tabular}{@{}lll}
\br
Quantity ($Q$) & Uncertainty ($\Delta Q/Q$) & Experiment\\
\mr
$R$ & $7.3 \times 10^{-9}$ & \cite{Andreas2011} \\
$M$ & $5.0 \times 10^{-9}$ & \cite{Andreas2011} \\
$T$ & $1.8 \times 10^{-8} $ & \cite{Lee2010} \\
$G$ & $6.3 \times 10^{-8}$ & \\
\br
\end{tabular}
\end{indented}
\end{table}

In \tref{tab.1}, we present preliminary error estimates using the fractional uncertainty of each relevant quantity sampled from various experiments. For a preliminary estimate, a period of the form in equation~\eref{SHOperiod} is assumed such that $G = 4\pi^2 R^3/MT^2$, where we approximate with $M$ as the total mass of the two-layered sphere and $R$ the outer radius (this expression differs from the actual period by about 4.69\% according to the numerical simulations below). Then
\begin{eqnarray}
\frac{\Delta G}{G} = 3\frac{\Delta R}{R} + \frac{\Delta M}{M} + 2\frac{\Delta T}{T},
\end{eqnarray}
is the rough fractional uncertainty of our $G$ determination. A more detailed equation will be given in future work, taking into account the two-layered nature of the sphere and the accuracy of producing the cylindrical hole. Note that mass inhomogeneities are not necessarily a major hurdle, as with previous $G$ experiments, considering the substantial advances in the fabrication of uniform spheres for the Avogadro Project \cite{Andreas2011, Becker2011}. Mass determinations of these spheres were made by comparing with the platinum-iridium alloy $1$~kg standards from Bureau International des Poids et Mesures (BIPM), National Metrology Institute of Japan (NMIJ) and Physikalisch-Technische Bundesanstalt (PTB). They are independent of the value of $G$ besides applied microgram corrections dependent upon little $g$ measurements at the calibration location of the employed balance \cite{Jabbour2001}.

\begin{figure}
\centering
\includegraphics[scale=.63]{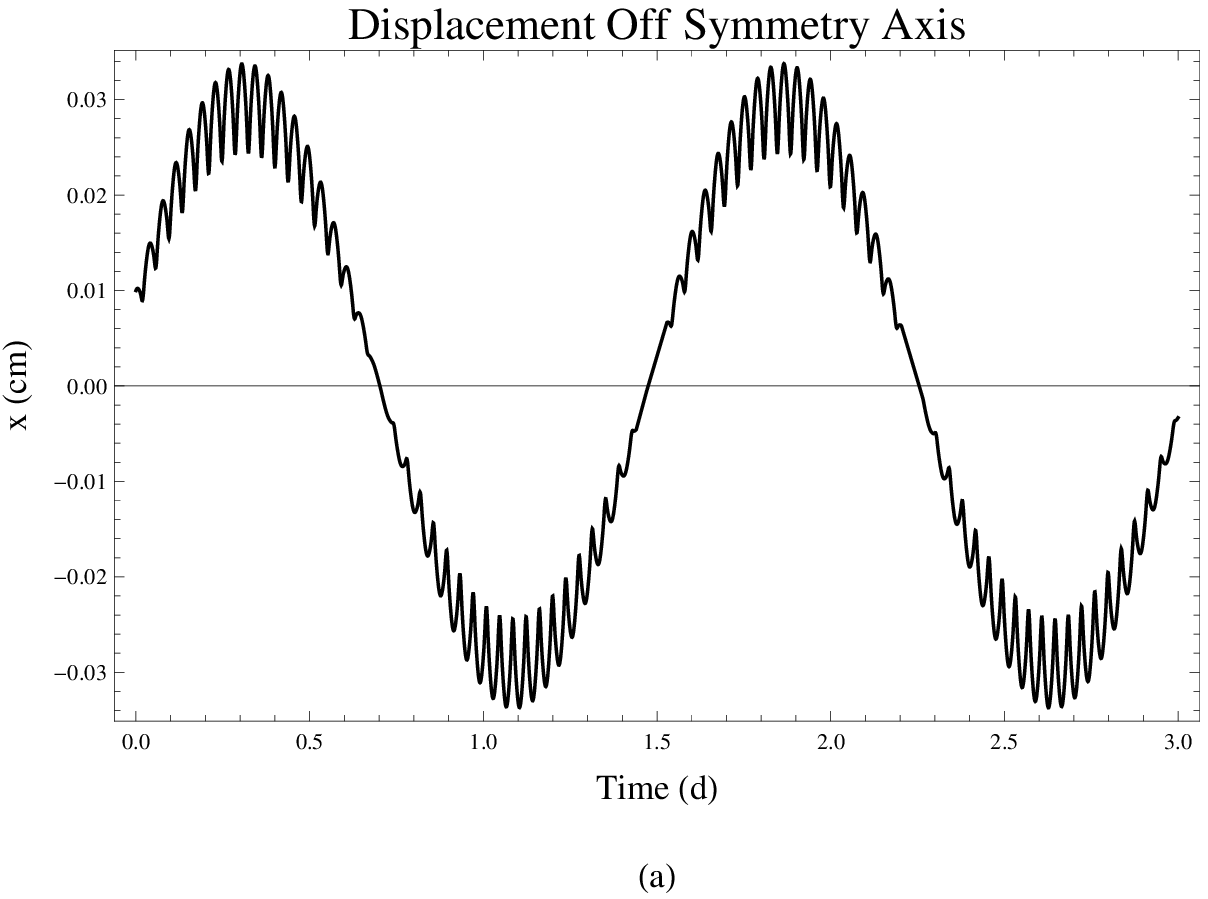}
\includegraphics[scale=.63]{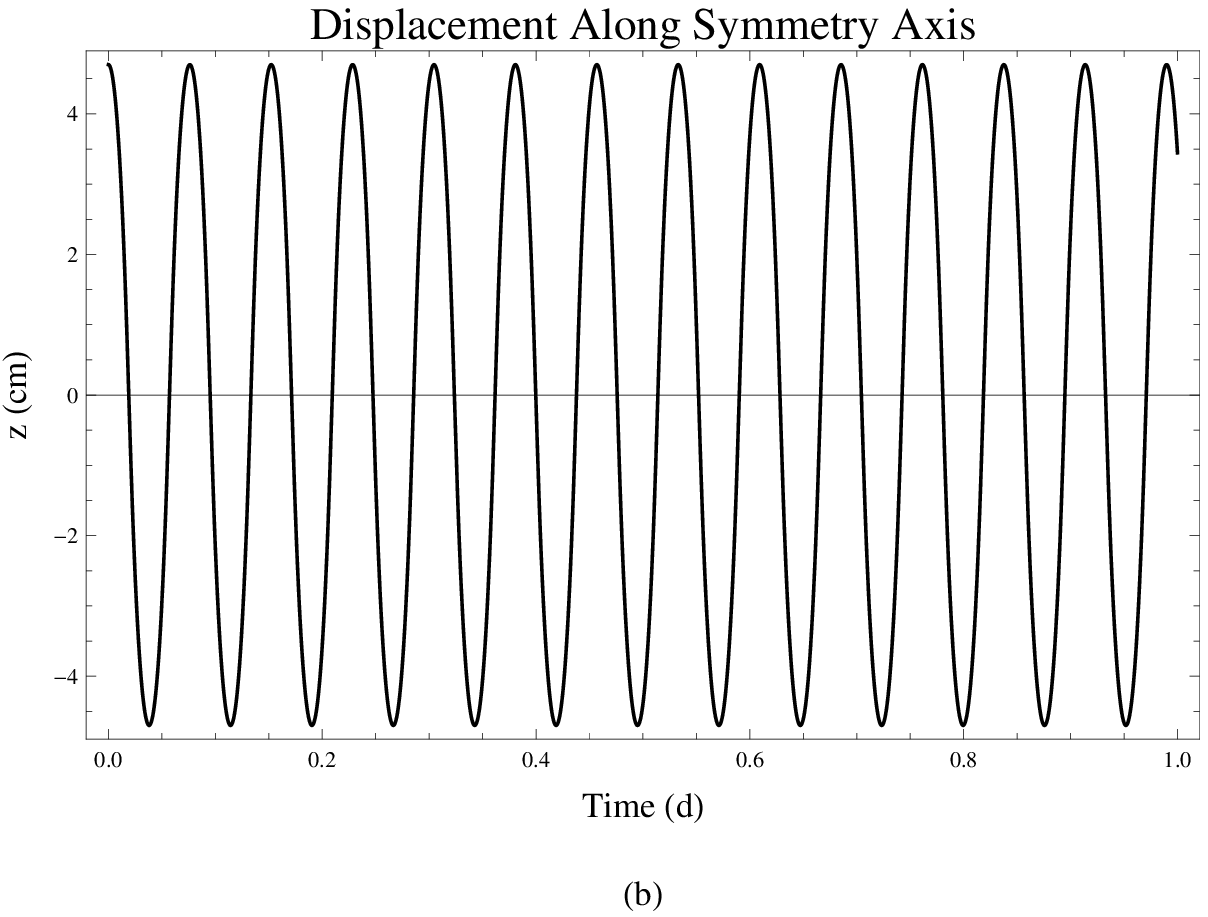} \\
\includegraphics[scale=.57]{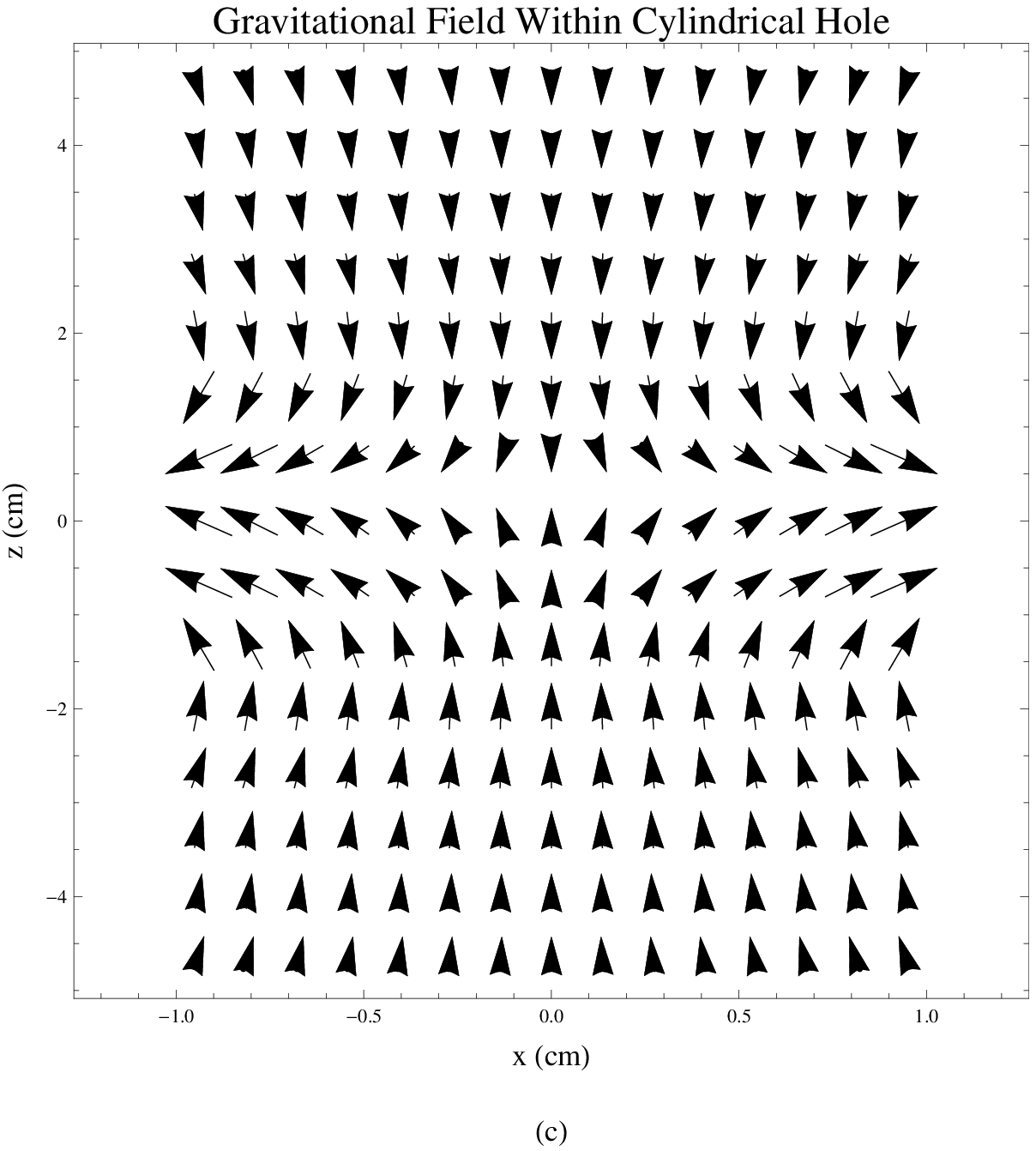}
\caption{Simulation of the test mass position over time under the presence of the gravitational field of the two-layered spherical ring, calculated to first order in $\varrho$. (a) Position in the plane perpendicular to the symmetry axis. (b) Position along the symmetry axis. (c) Vector field plot of the gravitational field inside the hole of the two-layered spherical ring. An inner core of platinum surrounded by an outer envelope of silicon is assumed, with parameters: $a = 0.94$~cm, $R_1 = 1.88$~cm, $R_2 = 4.7$~cm, $\mu_{1} = 21.45$~g~cm$^{-3}$, $\mu_{2} = 2.32$~g~cm$^{-3}$. The $z$-axis is aligned with the symmetry axis of the hole. The $x$-axis is perpendicular to $z$ and parallel to the initial transverse displacement of the test mass away from the symmetry axis (origin located at the center of the spherical ring). Initial conditions used: $z(0)=4.7$~cm, $\dot{z}(0)=0$, $x(0)=0.1$~mm, $\dot{x}(0)=1$~mm d$^{-1}$.}
\label{fig.3}
\end{figure}

We also have the advantage of determining positional measurements with proven ranging techniques \cite{Murphy2013}, which allows for a significant improvement in accuracy if one looks to recent applications of femtosecond laser pulses \cite{Lee2010, Lee2012, Lee2014}. To determine the fractional uncertainty in period, we employed the formalism of \cite{Kopeikin1999}, originally used for analyzing binary pulsar data. Paralleling equation~(52) of \cite{Kopeikin1999}, our uncertainty in period is taken to be
\begin{equation} \label{binaryPulse}
\frac{\Delta T}{T} = \sqrt{\frac{75}{4\pi^3 N^3}} \frac{\Delta z}{R},
\end{equation}
where $N$ is the number of oscillations used for each $G$ determination, $R$ is the radius of the larger sphere and $\Delta z$ is the range error for a single measurement. The fractional uncertainty in period from \tref{tab.1} is obtained by assuming the sphere dimensions of \cite{Andreas2011} for an outer radius of $4.7$~cm and the laser accuracy of \cite{Lee2010} for a range error of $1.1$~nm with a $G$ determination made once per oscillation. While the accuracy of the period measurement improves as the number of oscillations used to make a determination increases, there are issues with host spacecraft clock stability that must be taken into consideration since the period along the $z$-axis for these assumed values will be approximately two hours for a larger sphere mass of about one kilogram.

We perform a stability analysis with these assumed dimensions, taking the hole radius to be $a = 0.94$~cm, with an inner core of platinum, as an example, having radius $R_1 = 1.88$~cm, surrounded by the outer envelope of silicon (sphere total mass $\approx 1.3$~kg). From the numerical simulations of \fref{fig.3}, it is clear that stable periodic behavior in $\varrho$ is obtained due to the presence of the dense core. While the test mass oscillates between both ends of the tunnel along the $z$ symmetry axis in a harmonic manner with a period of 1.83~hr (\fref{fig.3}(b)), it also oscillates perpendicular to the symmetry axis with a period of about two days (\fref{fig.3}(a)), displaying a rich spectrum of shorter period oscillations as well. A preliminary simulation of one year of oscillator motion maintains this stable behavior, affirming the possibility of obtaining a significant amount of data with our apparatus. Note, however, the stringent initial conditions we've imposed for the velocity of the retroreflector relative to the sphere center after apparatus release from the host. Increasing the initial transverse velocity away from the symmetry axis and/or the initial displacement off the axis by a factor of ten preserves stability without impacting the wall. Maximum allowed initial transverse velocity is then $10$~mm~d$^{-1}$, almost two orders of magnitude smaller than the LISA Pathfinder release specifications. Similarly, the sphere-retroreflector system remains stable for initial velocities along the symmetry axis less than $1$~m~d$^{-1}$, to err on the conservative side.

There are certain issues with this setup which will be addressed in detail in subsequent work. Particularly, concern lies with the question of how to release the apparatus from the host spacecraft in a manner that ensures the stable initial configuration necessary to produce harmonic motion along the symmetry axis without the retroreflector bouncing between the walls of the tunnel, as briefly mentioned above. One would hope to constrain motion perpendicular to the symmetry axis of the hole with the appropriate initial conditions after release, which is certainly not a trivial task for a spinning spacecraft, although one could utilize the host spacecraft's angular momentum to spin the sphere about its symmetry axis, effectively averaging any mass inhomogeneities and likely reducing off-axis rotational instability (maximum off-axis rotational velocity of the sphere on the order of $10^{-9}$ rad s$^{-1}$ is needed to maintain laser pointing stability for at least one year). Rotational stability of the retroreflector itself after release from the host must also be accounted for, with the use of a spherical retroreflector as a possible solution to this issue, i.e. a miniature version of LAGEOS \cite{Cohen1985}. It will likely be necessary to resolve the spatial position and velocity of the retroreflector in the two-dimensional plane perpendicular to the symmetry axis, requiring a more detailed pulse detection mechanism onboard the host. Utilizing Doppler techniques in addition to range would be helpful for velocity resolution and ultimately provide greater flexibility in the analysis of obtained tracking data. Furthermore, one must boost the host away from the apparatus taking care not to significantly disturb its orientation by e.g. thruster plume impingement (our apparatus would likely be ``piggybacked" to deep space on a major mission as a minor portion of an extended mission phase). Detailed engineering considerations would be imperative to devise the appropriate release mechanism to attain initial displacements and velocities smaller than the maximum bounds mentioned at the end of the last paragraph.

All non-gravitational accelerations acting upon our two-body system must be minimized for an isolated environment (see table~II of \cite{Anderson2002} for a list). While this is the reason we suggest performing our experiment in deep space, solar radiation pressure is still relevant, even if significantly reduced. However, if we employ the shielding mechanism mentioned in previous deep space proposals \cite{Johann2008}, where the gravity oscillator remains in the shadow of the host spacecraft, this force is less of a concern on the sphere-retroflector apparatus as the host encounters all of the solar radiation. For a spacecraft antenna of $2.5$~m diameter at a distance of 25~AU and directed toward the Sun, one finds a shadow length of nearly $6.7$~km, which provides an informative upper bound for our spacecraft-apparatus distance. The momentum imparted by each laser pulse upon the target retroreflector will also need to be appropriately modeled as well as the reaction on the host of emitting and subsequently receiving such a pulse. The goal would be to find the optimal distance for our apparatus that minimizes the gravitational acceleration due to the host while maximizing the probability of successfully detecting a reflected range pulse with the minimum number of photons per pulse to keep the laser radiation pressure on the oscillator insignificant on the timescales used to determine each period measurement. The size and mass of the retroreflector will play a major role in the analysis (the technology is there for micron-sized retroreflectors \cite{Sherlock2011}, but these may not be massive enough to combat the laser radiation pressure). It may be necessary to model the laser radiation pressure if we are unable to decrease intensity below the threshold where our quoted $G$ uncertainty would be unaffected.

Besides also appropriately modeling thermal radiation emitted by radioisotope thermoelectric generators (RTGs) powering the likely spin-stabilized host spacecraft, one must consider sensitivity of the line-of-sight distance between the host and sphere-retroreflector apparatus to the gravitational attraction between the host and the sphere-retroreflector apparatus as well as to tidal effects produced by the influence of the Sun. For a radial trajectory with line-of-sight direction along the symmetry axis of the sphere, monopole effects due to, for instance, a 10~kg host produce an attractive acceleration on the apparatus of $1.5 \times 10^{-17}$~m~s$^{-2}$, if the sphere-retroreflector center of mass is taken to be $6.7$~km away. Additionally, solar tidal effects at a heliocentric distance of 25~AU for the same approximately $6.7$~km line-of-sight distance produce a relative acceleration of $3.4 \times 10^{-14}$~m~s$^{-2}$, separating the host from the apparatus. The next highest order term in the gravitational potential of the Sun ($J_2$) produces a tidal acceleration having magnitude of about $10^{-20}$~m~s$^{-2}$. Oversimplifying with a constant relative acceleration to determine order of magnitude effects on the line-of-sight distance, we find all relative accelerations between the host and the sphere-retroreflector center of mass greater than $5.1 \times 10^{-17}$~m~s$^{-2}$ must be modeled when analyzing the two-way pulse data, assuming one $G$ determination for each harmonic cycle. Accelerations greater than this limit produce a drift in the line-of-sight distance larger than $1.1$~nm for one $1.83$~hr period of the oscillator. Thus, only monopole solar tidal effects are relevant, but these are routinely modeled in orbit determination programs and will be easily accounted for in our line-of-sight determinations. Other forces on the host exhibit behavior distinct from that of the periodic signature of the gravity clock oscillator, allowing for a relatively simple extraction of the desired signal: for example, solar radiation pressure has $r^{-2}_{\odot}$ dependence and thermal from RTGs decays exponentially over time.

Stray small force gradients are a fundamental issue with $G$ tests. Our gravity clock geometry presents a stiffness (force gradient per unit mass of the test body) of $(2\pi/T)^2 = 9.1 \times 10^{-7}$~s$^{-2}$, which, using the $\Delta T/T$ uncertainty stated in \tref{tab.1}, requires eliminating or modeling extraneous stiffness greater than $3.3 \times 10^{-14}$~s$^{-2}$ for a 63 parts per billion (ppb) $G$ measurement resolution. Electrostatic patch fields \cite{Speake1996, Speake2003} produce stiffness measured to be much larger for LISA Pathfinder, on the order of $10^{-9}$~s$^{-2}$ \cite{Cavalleri2009}. Appropriate choices for the material of the retroreflector and inner core composition need to be carefully considered, and balancing a decrease in the size of the retroreflector with surface pressure forces, like laser radiation pressure, becomes important. Additionally, increasing the size of the hole helps to mitigate stray patch effects, but such an increase may introduce additional gravitational instabilities. The likely approach forward is a detailed study of possible materials to be used with the mindset of appropriately modeling electrostatic patch field effects in the orbit determination of the test mass. One must also consider stiffness due to solar tidal accelerations between the oscillating retroreflector and the center of mass of the sphere. However, assuming the line-of-sight direction along the symmetry axis coincides with the radial direction away from the Sun, the first order solar tidal term on the apparatus itself at 25~AU adds a stiffness of $2M_{\odot}G/r_{\odot}^3 = 5.1 \times 10^{-18}$~s$^{-2}$, which is negligible for our proposed sensitivity.

Similarly, charging of the oscillator due to cosmic ray interactions \cite{Araujo2005} contributes another force gradient within the hole. Accumulated charge on the test mass eventually results in an unstable system as the oscillator is attracted toward its image charge on the wall of the tunnel (similar issues arise from charge accumulation by the solid sphere). For a charge management system, one could take the approach of \cite{Buscaino2015} and shield the apparatus within a spherical conductor shell ($\sim$ 1~m in radius). Intermittent exposure of the apparatus to ultraviolet radiation from light-emitting diodes on the interior lining of the shell provides a discharge mechanism for the gravity clock. An opening to the interior of the shell from the perspective of the host spacecraft is, however, needed to preserve the laser line-of-sight to the oscillator. Such a shell would be equipped with thrusters to rotate and maintain this viewing window orientation (alternatively, a relay system as in \cite{Buscaino2015} would also work). An active shielding mechanism, such as ``mini-magnetospheres" \cite{NASATM2005, Bamford2014}, provides a complementary approach to combating charge accumulation, yet this technology currently needs to mature significantly. Any potential gravitational perturbations introduced due to the presence of an enclosing shell must be taken into consideration as well. Nevertheless, given these issues with the release mechanism, laser radiation pressure and stray force gradients, the uncertainty stated in \tref{tab.1} for $T$ is likely optimistic, and our estimate of $63$~ppb for the accuracy of this $G$ determination seems to be a lower bound.

Lastly, we look to any possible relativistic effects. Taking the idealistic scenario of the oscillator moving within a spherically symmetric perfect fluid of radius $R$ with constant density and mass $M$, the spacetime metric assumes the form (see equation~(6) of box~23.2 in \cite{Misner1973})
\begin{equation}
ds^2 = -\frac{c^2}{4}\bigg[3\sqrt{1-\frac{2MG}{c^2R}} - \sqrt{1-\frac{2MGr^2}{c^2R^3}} \bigg]^2 dt^2 + \frac{dr^2}{1-\frac{2MGr^2}{c^2R^3}} + r^2 d\Omega^2.
\end{equation}
Examining geodesic motion, the coordinate acceleration of the radially traversing oscillator released from the fluid surface as seen by a fiducial observer at infinity is found to be
\begin{equation}
\frac{d^2 r}{dt^2} = - \bigg[\bigg(\frac{MG}{R^3}\bigg) - \frac{3}{2c^2}\bigg(\frac{MG}{R^2}\bigg)^2 \bigg]r + \frac{1}{2c^2} \bigg(\frac{MG}{R^3}\bigg)^2 r^3 + \mathcal{O}\bigg(\frac{1}{c^4}\bigg) .
\end{equation}
Time measured in terms of the clock of a fiducial observer at infinity is taken here as a valid approximation for the time measured by a clock onboard the host spacecraft. In terms of $r/R$, relativistic correction terms are proportional to $(MG/cR^2)^2$, rendering them unimportant given the sensitivity of our experiment, as expected. Interestingly, under the assumption the oscillator is released from rest near the fluid center, i.e. $(r/R)^3$ term $\approx 0$, one recovers simple harmonic motion but with period
\begin{equation}
T = 2\pi \sqrt{\frac{R^3}{MG}} \bigg[1+\frac{3}{4}\frac{MG}{c^2R}\bigg],
\end{equation}
which, for an Earth-sized mass, amounts to an additional $2.6$~$\mu$s according to the clock of the fiducial observer at infinity.

\section{Conclusions}

We've offered a novel experiment to measure Newton's gravitational constant $G$. Employing our gravity clock in deep space in cooperation with ranging capabilities on a host spacecraft, it seems possible our apparatus could attain a fractional uncertainty in measuring $G$ at the level of $6.3 \times 10^{-8}$ using existing technology, a potential improvement of nearly three orders of magnitude from current Earth-based measurements \cite{Mohr2015}. One has the added benefit with this proposal of further testing our understanding of gravitation by searching for minute accelerations acting upon the host spacecraft. Still, significant technological advances for the release mechanism from the host are likely required to ensure stable initial conditions for the apparatus. An active release mechanism which could reorient the sphere-retroreflector setup if the desired initial conditions were not met would be ideal. Accurate modeling of stray force gradients and laser radiation pressure along with optimizing the dimensions and materials of the apparatus geometry are outstanding issues for a more detailed study. Engineering considerations will also be provided in subsequent work.

Our approach to measuring $G$ also highlights the interesting mathematical problem of determining the motion of a test particle in vacuum but surrounded by a particular mass distribution, important for experimental gravitational physics. Extending the first order in $\varrho$ calculations displayed in \fref{fig.3}(c), one determines an integral expression for the potential inside the hole of a single layer homogeneous sphere,
\begin{eqnarray}
\fl \psi (\varrho, z) = -2\pi G\mu R^2 \sum_{j=0}^{\infty} \bigg[\frac{(2j)!}{4^j (j!)^2}\bigg]^2 \bigg(\frac{\varrho}{R} \bigg)^{2j} \nonumber
\\ \times \int_{-\sqrt{1-(\frac{a}{R})^2}}^{\sqrt{1-(\frac{a}{R})^2}} dx \, \frac{x (x - \frac{z}{R})}{(1-x^2)^{j+1/2}} F\bigg(j+\frac{1}{2}, j+\frac{1}{2}; \frac{3}{2}; - \frac{(x-\frac{z}{R})^2}{1-x^2} \bigg),
\end{eqnarray}
using the Green's function in cylindrical coordinates \cite{Arfken1985}, a $\varrho$ power series expansion of the zeroth order modified Bessel function of the first kind and equation~(3) in section 6.699 of \cite{Gradshteyn1994}, after appropriate integration by parts. $F(a,b;c;z) = {}_2 F_1(a,b;c;z)$ is the hypergeometric function. Of course, the principle of superposition can then be applied to obtain the potential for the multilayered object. Mathematica appears able to calculate the relevant indefinite integral expressions to at least $\mathcal{O}(\varrho^{20})$ \cite{Mathematica} and verifies the zero and second order terms of \eref{gravPotentialLowOrder}. Determining the gravitational field if the hole is slightly displaced away from the diameter of the sphere is an individual example of other modeling issues that will need to be addressed.

Furthermore, if it were somehow possible to also range to the larger sphere with a second pulse, we'd have an interesting test of the equivalence principle in deep space. One could compare the rate of free fall of the larger sphere toward the host versus that of the smaller retroreflector after accounting for the motion of all three objects in the Solar System barycenter (SSB) inertial frame, although any detection may be difficult to differentiate from a variation in the oscillator's period and thus in $G$. The accuracy of such a test would depend on knowledge of the smaller retroreflector mass as well as the motion of the host.

\ack
We thank the referees for helpful comments.

\section*{References}
\bibliography{gravtrainCQG}

\end{document}